# Logical segmentation for article extraction in digitized old newspapers

Thomas Palfray
LITIS, University of Rouen
UFR des Sciences et Techniques
F-76800 Saint Etienne du Rouvray

thomas.palfray@univ-rouen.fr

David Hébert
LITIS, University of Rouen
UFR des Sciences et Techniques
F-76800 Saint Etienne du Rouvray

david.hebert@univ-rouen.fr

Stéphane Nicolas
LITIS, University of Rouen
UFR des Sciences et Techniques
F-76800 Saint Etienne du Rouvray

stephane.nicolas@univ-rouen.fr

Pierrick Tranouez
LITIS, University of Rouen
UFR des Sciences et Techniques
F-76800 Saint Etienne du Rouvray

pierrick.tranouez@univ-rouen.fr

Thierry Paquet
LITIS, University of Rouen
UFR des Sciences et Techniques
F-76800 Saint Etienne du Rouvray

thierry.paquet@univ-rouen.fr

## ABSTRACT
Newspapers are documents made of news item and informative articles. They are not meant to be red iteratively: the reader can pick his items in any order he fancies. Ignoring this structural property, most digitized newspaper archives only offer access by issue or at best by page to their content.

We have built a digitization workflow that automatically extracts newspaper articles from images, which allows indexing and retrieval of information at the article level. Our back-end system extracts the logical structure of the page to produce the informative units: the articles. Each image is labelled at the pixel level, through a machine learning based method, then the page logical structure is constructed up from there by the detection of structuring entities such as horizontal and vertical separators, titles and text lines. This logical structure is stored in a METS wrapper associated to the ALTO file produced by the system including the OCRed text. Our front-end system provides a web high definition visualisation of images, textual indexing and retrieval facilities, searching and reading at the article level. Articles transcriptions can be collaboratively corrected, which as a consequence allows for better indexing.

We are currently testing our system on the archives of the Journal de Rouen, one of France eldest local newspaper. These 250 years of publication amount to 300 000 pages of very variable image quality and layout complexity. Test year 1808 can be consulted at plair.univ-rouen.fr.

## Categories and Subject Descriptors
I.7.5 [**Document and Text Processing**]: Document Capture – *Document analysis*

## General Terms
Algorithms.

## 1. INTRODUCTION
During the last twenty years, the archives and national libraries of the whole world implemented many programmes of digitalization of their patrimonial funds in order to preserve them while facilitating the access of the public to information they contain. The case of old newspapers is emblematic of this will of historical dissemination of information, because of the wealth and the diversity of the contained information in the documents of this type. Nevertheless, these documents require particular treatments to fully exploit their informational contents, because of their size, of their particularly complex page layout and its evolution with the technical innovations of printing works,. Moreover, the quality of conservation of this kind of document is often variable depending on the historical periods because of the important variations of quality of paper used, resulting sometimes in a very degraded digitized copy, even unusable. In order to exploit these documents as well as possible, it is necessary to have a segmentation in articles making it possible to isolate the interesting parts of a newspaper for an easier consultation by the user, by the means of modern digital tools. Having in mind these difficulties, we developed a new method for logical labelling of old newspapers. This method is intended to extract metadata in the images of the digitized newspapers, thanks to the joint use of a method of classification of sequence of pixels based on conditional random field modelling, associated with a set of rules defining the concept of article within a newspaper. In a first time we describe the previous work related to this task in the literature, then we fully describe the method we propose before presenting the results obtained on newspaper issues from "Journal de Rouen", a regional French newspaper. Finally we describe briefly the complete process in which this method is integrated, and then we conclude by a discussion about the potential of the method and future work.

## 2. Related work
Since 2001 is organised in the context of the ICDAR Conference, a document page segmentation competition [1] in which some of the proposed algorithms may have goals similar to the system we propose in this paper. Nevertheless, the document dataset used for this competition contains only modern documents; therefore the proposed methods may be inefficient for old newspapers. We can cite however the work described in [2] which exploits a method of labelling at pixel level adaptable to this type of documents. [3] describes an approach based on the determination of the maximal empty rectangles to delimit columns and text blocks. This method is integrated in the system OCROPUS. Although interesting, this method does not allow for difficulties inherent to old documents (skewing, deformations,...). A more interesting method taking into account these difficulties is described in [4]. The authors propose to use a multiscale approach to extract text blocks in old newspapers. This method seems to be efficient, but as the previous one, only the segmentation in text blocks is provided, but no logical reading order, what is important to determine the logical structure of articles. The approach we propose in this paper tries to bring a solution to this problem.

## 3. Proposed approach

Binary images of the digitized newspaper issues are considered at the input of the system, and files in METS format are produced at the output. The METS files contain the structuration of the newspaper issues in articles, and the ordering of the articles according to the reading order. To do that we want to identify the editorial model of the document using visual clues determined thanks to a conditional random field model associated to a structural analysis using a recursive algorithm. The distinctive steps of the proposed method are the following ones:

- logical labelling at pixel level using conditional random field modelling
- smoothing of the labelling using majority vote
- extraction of the text lines
- generation of a segmentation mask using identified rulers
- recursive analysis of the document page to extract articles and determine the reading order

## 3.1 Logical labelling at pixel level

The proposed method for article extraction in newspaper document images relies on a segmentation stage using an analysis of the document image by a particular conditional random field (CRF) model with multi-scale quantization feature functions. This system provides a fine segmentation at pixel level, where each pixel is associated to a logical label specifying the logical function of the entity this pixel belongs to. The segmentation obtained thus does not consist only of one division in physical blocks but it also brings a logical identification to the detected entities. See [5] for more details about this model. This segmentation considers ten logical labels describing precisely the physical organisation of the textual content, and particularly the inter and intra character spacing. These ten labels are finally grouped into six informative labels for the logical labelling into articles or article parts:

- vertical separator
- horizontal separator
- title (compound by the labels "title character", "title inter-character" and "title inter-word")
- text line (compound by the labels "character", "inter-character" and "inter-word")
- noise
- background

Each image is analysed line by line at the pixel level, and the labelling results of all the lines are concatenated to produce a labelled image noted $Iseg$, which is then exploited by the next analysis stages. An example of labelling obtained at this stage is shown on Figure 1b. At the end of this stage we dispose of a set of elements identifying the vertical and horizontal separators, the titles, and the text lines. However these elements are locally detected and they might be fragmented or erroneous. A complementary analysis using grouping rules defined by a simple and generic layout model is then applied to analyse the spatial placing of these entities and thus allow to the extract the articles.

## 3.2 Post processing for the logical extraction of the articles

### 3.2.1 Labelling smoothing using majority vote

Despite the ability of the CRF model to label each region or pixel according to the underlying layout model, some labelling error may appear. For this reason we apply a first algorithm to smooth the labelling result by applying a majority vote on the labels associated to the black pixels of the binary image (which correspond to the informative pixels by comparison with the white ones which correspond generally to the background), as described by algorithm1. The results provided by this smoothing algorithm are shown on figure 1c.

**Algorithm 1**: majority vote on connected components

**input**: $Iseg$, image resulting of the segmentation
**output**: $Ientity$, image with the extracted entity

Extract in $Iseg$ the connected components constituted by all the connected labels except those of the "background" label;
**For all** the connected components CC **do**
$nEi \leftarrow$ the number of pixels labelled $Ei$;
Affect to all the pixels of CC, the label $argmax_{Ei}(nEi)$

Once the image $Ientity$ is entirely built, we can extract the text lines and the articles.

### 3.2.2 Text line extraction

The text lines are obtained by extracting the connected components, which are labelled "text" in the resulting image $Icc$. In spite of the robustness of our extraction method, some text lines may be connected because of important deformation in the image due to document degradations or digitization artefacts. To solve this problem we try to detect the connected text lines by calculating the mean surface of the text lines, contained in a newspaper issue. We pose the assumption that the text lines, whose convex hulls delimit a surface much higher than the mean surface, are certainly erroneous. These text lines are then corrected by a specific algorithm, which allows separating them, before being associated to the article blocks extracted using our segmentation method

### 3.2.3 Article extraction using a layout model

#### 3.2.3.1 Article definition

As we said before we define on the image some area of interest representing the following entities: title, text, horizontal separator, and vertical separator. The definition of an article follows some precise layout rules, this is why we consider that an article begins with a "title" entity followed by at least one "text" entity, and ends by a "horizontal separator" entity or another "title" entity. This basic layout model also has to take into consideration the particular case of the articles, which extend on several pages. We consider that a "text" entity placed in the first or in the last position of a list of articles is linked to an article of the previous or the next page of the considered newspaper issue.

### 3.2.4 Definition of the grid of separators

The horizontal and vertical separators of the newspaper pages constitute robust information to identify the structure of the pages and extract the articles, but these separators are sometimes broken due to the degradations of the document. The vertical separators delimit the columns while the horizontal ones separate either the articles belonging to a column or the different sections of the page. The title areas are also important for the analysis of the logical structure of the page because they allow locating the beginning of an article. Thus the document being analysed can be modelled as a tree of blocks, each one of the blocks representing the different sections of this document. The root of this tree is constituted of one block representing the whole page, as the next levels are constituted by the blocks delimited by the vertical and horizontal separators, and so on. We call hierarchical position, the position of a block or of a separator in the tree representing the structure of the document. The set of all the separators of a given page constitute a logical grid describing the logical structure of the document (figure 1d). We exploit this information to define a list of articles logically arranged and ordered according to the reading order determined both by their position in the document and their hierarchical position in the tree describing the logical structure. Finally one or several boxes of the logical grid compose each one of the extracted articles.

### 3.2.4.1 Generation of the grid and text blocks extraction

The first step of our segmentation method consists in prolonging all the area of interest labelled as "separator" or "title" in order to generate a grid of separators representing all the articles of the document. For that we apply the following algorithm:

- Create the vertical and horizontal separator mask
- Connect the closed vertical separator
- Prolong the vertical separators as long as they do not cross a horizontal separator or a title
- Connect the closed horizontal separator
- Prolong the horizontal separators and the titles as long as they do not cross a vertical separator

Thus we obtain a full grid covering the entire image and then we can use it to extract the articles. For that we generate the list of blocks which correspond to the boxes of our grid and we compare the coordinates of these blocks to the coordinates of the text lines extracted by the procedure explained in the subsection 3.2.2. We extract also the "title" entities and we add to the blocks the text lines which are fully included. The blocks containing no text line are rejected from the list. Finally we obtain a list of boxes positioned on the separator grid, and we only have to use this list of boxes to obtain a list of articles ordered according to the reading order of the document defined by the editorial model. The figure 6 shows an example of reading order interpretation.

### 3.2.4.2 Article extraction and reading order determination

The boxes of the previously described grid may contain a full article or part of article. By part of article we mean a textual zone which is broken into several parts and covers several section or pages. To be able to regroup these parts of a same article, it is necessary to identify the reading order in the section including them.

As explained in the subsection 3.2.3.2, the separators divide the page in sections, the sections into subsections and so on up to the block level which corresponds to the leaves of the logical tree model. In the subsection 3.2.3.3 we explained how the boxes of the grid are determined, and we now explain how we determine the sections.

For each section $s$, we seek the nearest horizontal separator strictly longer than the considered section and situated just above it. This horizontal separator delimits two sections. Among them the one containing $s$ is called the parent section. We repeat this process until having considered the entire tree model from the leaves to the root.

In each intermediate node of the tree, we order the contained children sections. We consider that a section precedes another if its top-left coordinate is strictly on the left or higher in the document image. For the children sections corresponding to the leaves of the tree model, this order relation implies the reading order of these sections. If a longer separator delimits a box which contains no title, it means that this box is the second part of a fragmented article. In this case this part of article is grouped with the box just before it.

Finally at this end of this stage we obtain the list of articles ordered by sections.

## 4. Results

This method was tested on a dataset containing 42 document images issued from a French regional newspaper called "Journal de Rouen". The results are determined visually because we do not have the ground-truth data to check them automatically. We determined the detection rate and the over-segmentation rate. These results are given in Table 1.

**Table 1. Results of the logical segmentation into articles**

| #articles | #detected | #correct | %correct | %over-seg |
|---|---|---|---|---|
| 226 | 245 | 194 | 85.84 | 8.41 |

The analysis of the errors produced by our method allow us to see that a great number of them are due to the labelling errors at the output of segmentation stage performed by the CRF model. For example on our test dataset we have 14 text lines labelled by error as "title" causing the creation of a new article when the editorial rules described in subsection 3.2.3.1 are applied, and then we obtain 28 articles instead of the 14 we should find. Theses errors cause in general an over-segmentation.

## 5. Full processing system

The method we present in this paper is integrated in a full system able to process a large amount of old newspaper document images. This system first applies on the images binarization and skew correction processing before applying the article extraction method described previously. Then the extracted text lines are used to feed on OCR engine to recognize the textual content. Our system provides as output XML files in METS/ALTO format. These files contain both the logical structure describing the reading order of the articles and the physical layout constituted by the detected text lines and the associated OCR results. They can be used to index the documents to improve the online access to the data they contain in digital archive applications. For that we propose a web newspaper browsing application, which exploits the data produced by the system presented in this paper. This front-end system, which will be presented in a further paper, provides a web high definition visualisation of images, textual indexing and retrieval facilities, searching and reading at the article level. Articles transcriptions can be collaboratively corrected, which as a consequence allows for better indexing. We are currently testing the system on the archives of the Journal de Rouen, one of France eldest local newspaper. These 250 years of publication amount to 300 000 pages of very variable image quality and layout complexity. Test year 1808 can be consulted at plair.univ-rouen.fr.

## 6. Conclusion and future work

We presented in this paper a logical segmentation method based on the analysis of low-level labeling results produced by a CRF model, using a set of grouping rules defined by a generic layout model. The proposed method is able to segment the textual content of old newspapers with complex Manhattan structure (multi columns), using a little set of simple rules. We obtain with this method an article segmentation rate of 85.84% on a test dataset containing 42 images of "Journal de Rouen", one of the oldest French regional newspapers.

These first results are promising, and allow us to identify the main improvement issues. In a future work we will improve further both the CRF model and the layout rules in order to be able to take into account some important other entities of the document structure, such as figures, pictures, captions and tables.

## 7. ACKNOWLEDGMENTS

The PlaiR project (Regional Indexing Platform, in French "Plateforme d'Indexation Regionale") is funded by the "Haute-Normandie" regional council and the European institutions by the FEDER program. This project is supported by the CHU and the University of Rouen.

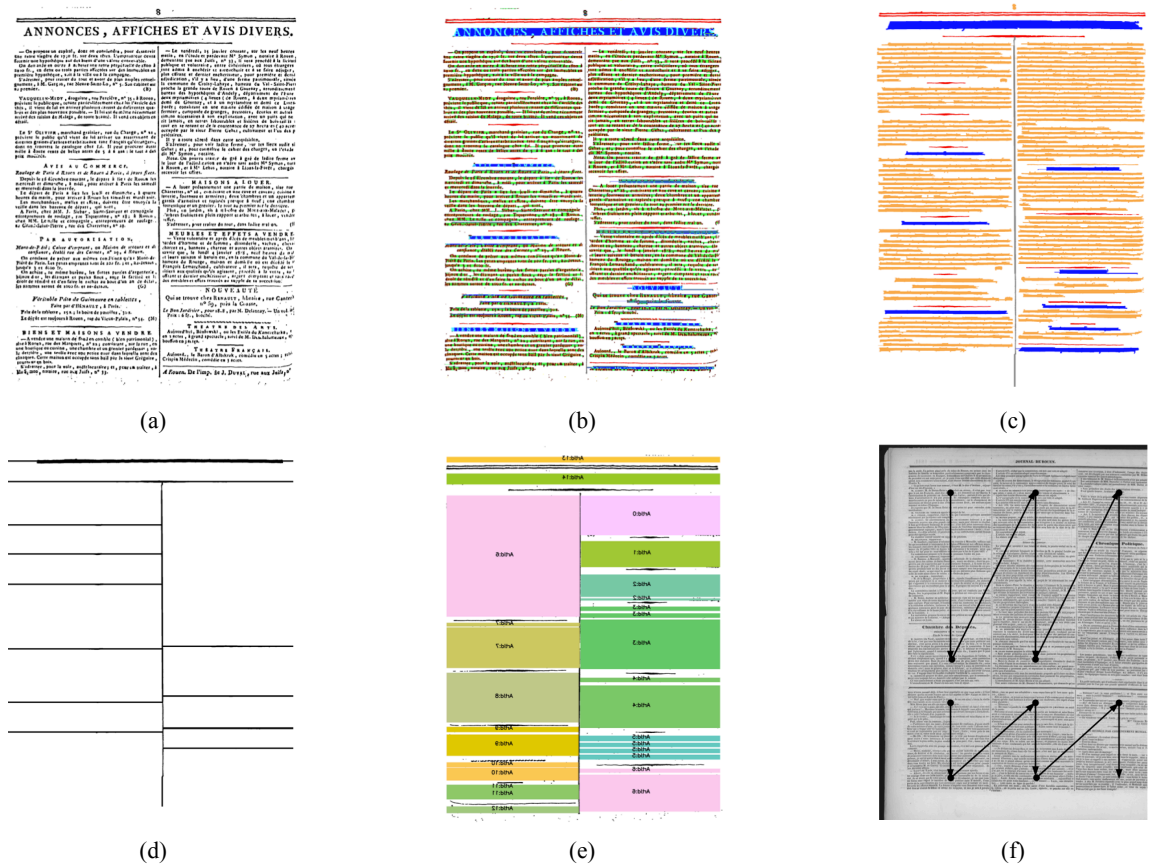

**Figure 1. (a) original image (b) pixel level logical labeling result (c) labelling smoothing (d) grid of separators (e) result of the segmentation into articles (f) example of reading order**